\documentstyle[12pt]{article}
\begin{document}
\newcommand{\be}{\begin{eqnarray}}
\newcommand{\ee}{\end{eqnarray}}
\newcommand{\zt}{\zeta}
\newcommand{\ve}{\epsilon}
\newcommand{\al}{\alpha}
\newcommand{\gm}{\gamma}
\newcommand{\bt}{\beta}
\newcommand{\dt}{\delta}
\newcommand{\la}{\lambda}
\newcommand{\vp}{\varphi}
\newcommand{\nn}{\nonumber}
\begin{center}
 { \Large\bf
 New approach to summation of field-theoretical \\
 series in models with strong coupling

 }
\bigskip
 {\Large A. I. Mudrov$^*$,  K. B. Varnashev$^{**}$}

\smallskip

{\it
$^{*}$
Institute of Physics, St.Petersburg State University, Ulyanovskaya 1,
Stary Petergof, St. Petersburg, 198904, Russia,\\
aimudrov@dg2062.spb.edu\\
$^{**}$
State Electrotechnical University,
Prof. Popov Street  5, St.Petersburg, 197376, Russia\\
feop@eltech.ru
}
\end{center}
\begin{abstract}
A new approach to summation of divergent field-theoretical series is
suggested. It is based on the Borel transformation combined with a
conformal mapping and does not imply the knowledge of the exact asymptotic
parameters. The method is tested on functions expanded in their
asymptotic power series and applied to estimating the
ground state energy of simple quantum mechanical problems
including anisotropic oscillators and caclulating the critical exponents
for certain comformal field models. It can be expected that the
new approach to summation may be used to obtaining numerical
estimates for important physical quantities represented by divergent
series in two- and three-dimensional field models.

\end{abstract}

\newpage

In this brief report we discuss the problem of summation of "bad"
series arising in various fields of physics \cite{1,2}.
By the strong coupling we mean the situation when the parameter
of expansion does not belong to the range of convergence of the
perturbative series. The typical situation is when the radius of
convergence is just zero and the resulting series are asymptotic.
On the other hand, these series can represent important physical 
quantities, and to extract reliable  information from them,
they should be processed by a proper resummation technique.

The main goal of the present work is to suggest a new approach to
treating divergent series and to apply it to finding numerical estimates
of the critical exponents for two- and three dimensional anisotropic
field models. The resummation technique proposed is tested on simple
model functions expanded in their asymptotic power series,
estimating ground state energy of isotropic and cubic anharmonic
oscillators, Yukawa potential, and critical exponents for conformal
field theories.

At present, there are several resummation procedures such as
simple Pad$\acute {{\rm e}}$, Pad$\acute {{\rm e}}$-Borel, 
and Pad$\acute {{\rm e}}$-Borel-Leroy techniques, whose application, 
however, is limited to series with coefficients alternating in signs.
For the more sophisticated method based on the Borel transformation 
combined with a conformal mapping, first proposed in Refs. \cite{3,4}, 
this limitation is not crucial. This technique 
was then elaborated and systematically used for various phase
transition problems \cite{5,6,7,8,8a,9,10,11,11a,12} being now   
regarded to as a most universal procedure. 
But it requires the knowledge the exact 
asymptotic high-order behavior of the series.
As a rule the coefficients of the series $\sum_k f_k g^k $
behave at large $k$ as $k! k^{b_0} (- a_0)^k$. The numbers
 $a_0$ and $b_0$ characterize the main divergent part.
Nowadays these parameters are found only for the simplest
case of the $O(N)$-symmetric models \cite{13,13a,14}, and
calculating them for anisotropic models is a most difficult problem
as yet unsolved. As an exception we mention the anisotropic quartic
quantum oscillator which represents
a one-dimensional $\varphi^4$ field theory with the cubic
anisotropy. For the perturbation expansion of the ground state energy
of this system the asymptotic parameters were found in Ref. \cite{15}.
Within the assumption of the  weak anisotropy the large-order asymptotic
behavior of the $\beta$-functions for the cubic model  was
deduced in Ref. \cite{16} and then used for determination of the
stability of the cubic fixed point in three dimensions \cite{17}.

Below, we make an attempt to overcome the outlined difficulties and
suggest a new approach to summation of the divergent field-theoretical
series, which is based on the standard technique Borel transformation 
combined with a conformal mapping \cite{1,2}, but which does not involve 
the exact values of the asymptotic parameters. We start from the 
Borel-Leroy transformation modified with a conformal mapping in 
the form \cite{7}
\be
 F(g;a,b) =\sum_{k=0}^\infty A_k(\la) \int_0^\infty
 e^{-\frac{x}{a g}} \Bigl(\frac{x}{a g}\Bigr)^b
 d\Bigl(\frac{x}{a g}\Bigr)
 \frac{\omega^k(x)}{(1-\omega^k(x))^{2\la}}.
 \label{eq1}
\ee
The coefficients $A_k(\la)$ are determined from the equality
$B(x(\omega))=\frac{A(\la,\omega)}{(1-\omega)^{2\la}}$
where
$\omega=\frac{\sqrt{x+1}-1}{\sqrt{x+1}+1}$
and the Borel transform $B(x)$ is the analytical continuation
of the series $\sum_k \frac{f_k}{a^k\Gamma(b+k+1)} x^k$ absolutely
convergent in the unit circle, $f_k$ being the coefficients of the
original series. Parameter $\la$ is choosen from the condition
of the most rapid convergence of series (\ref{eq1}),
 that is from minimizing the quantity
 $|1-\frac{F_L(g;a,b)}{F_{L-1}(g;a,b)}|$, where $L$ is the step
 of truncation and $F_L(g;a,b)$ is the $L$-partial sum for $F(g;a,b)$.
In the regular scheme \cite{4,7,8} parameters $a$ and $b$
are related to the exact asymptotic values $a_0$ and $b_0$. Since,
in practice we deal with a piece of the series only, where the
asymptotic regime might not be established, we vary parameters 
$a$ and $b$ in a neighbourhood of their exact values \cite{18}. 
Our principle observation
is that the result of processing $F_L(g;a,b)$ exhibit very
weak dependence on the transformation parameters $a$ and $b$ varying
in a wide range. The dependence becomes weaker with the growth
of the approximation order and the smaller is the parameter of expansion 
$g$ the better this property holds. We put the stability of the result of
processing with respect to variation of $a$ and $b$ into the foundation
of our technique to summation of divergent series. Such an approach allows
us to apply the transformation (\ref{eq1}) even if the exact asymptotic
behaviour of the series being processed is unknown.

The detailed analysis of the formulated resummation scheme
applyed to the model functions
 $${\cal F}(g) = \int_{-\infty}^{+\infty} e^{-x^2-g x^4}dx
 \sim\sum_{k=0}^\infty (-1)^k \frac{\Gamma(2k+\frac{1}{2})}{k!}g^k,$$
 $${\cal E}(g) =\int_0^\infty e^{-x}(x\partial_x)^{b_0}\frac{1}{1+g x}dx 
 \sim
 \sum_{k=0}^\infty (-1)^k k! k^{b_0} g^k $$
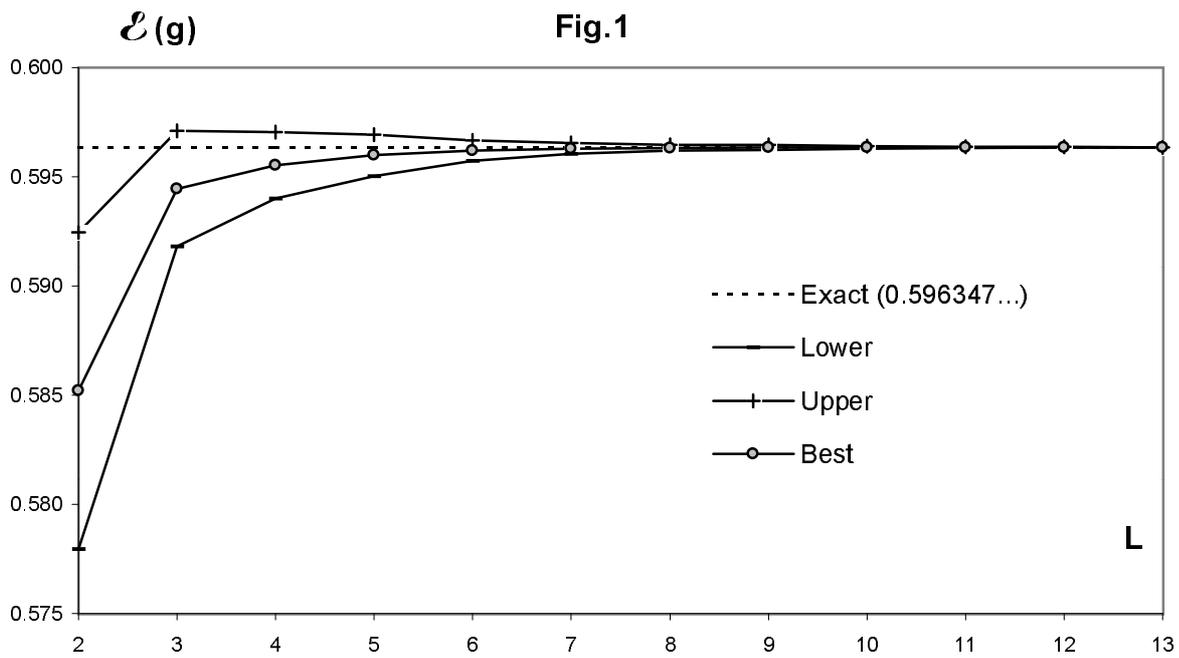
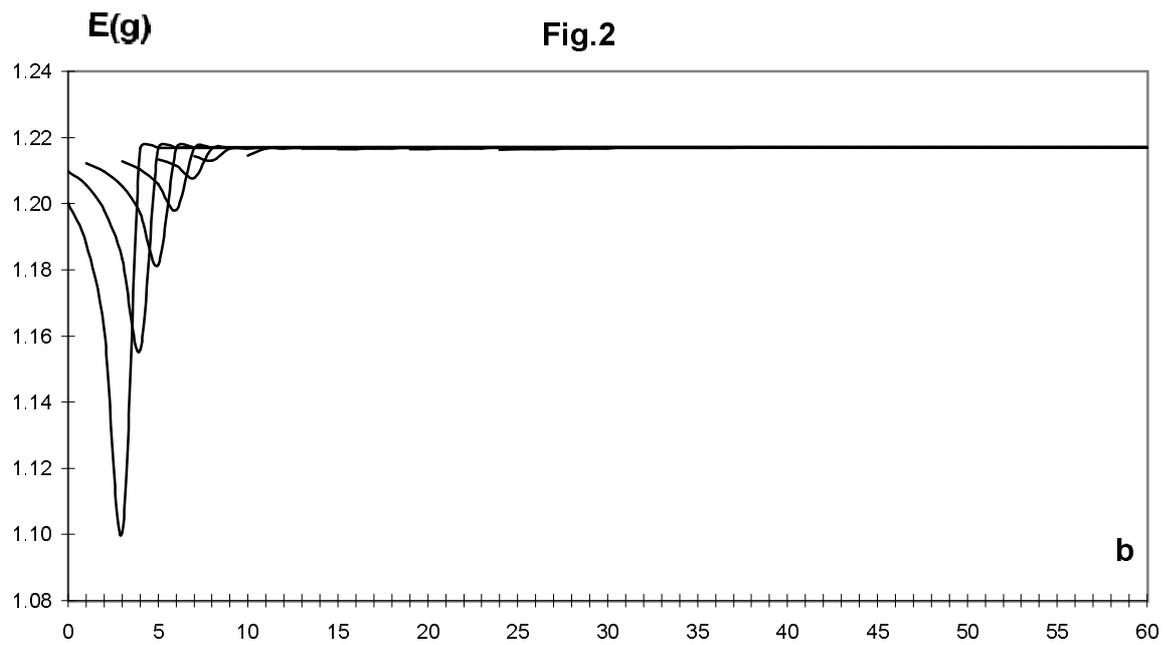
was performed in Ref. \cite{19}. The convergence of the process for the
function ${\cal E}(g)$, $g=1$ is presented in Fig.1.
Let us demonstrate our summation approach by estimating the ground
state energy $E(g)$ for several simple quantum mechanical systems.
Consider first the isotropic anharmonic oscillator \cite{20} with the
Hamiltonian $H=x^2+g x^4$. We observe the same stability of the result
of processing $E(g)$ with respect to $a$ and $b$
as for the model functions. The ground state energy estimates
for $g=1$ depending on the approximation order are listed in Table I.

\vspace{0.5cm}
 TABLE I. Numerical estimates for the isotropic anharmonic oscillator
 ground state energy at $g=1$.

\vspace{0.3cm}
\hspace{0.2cm}
\begin{tabular}{|c|c|c|c|c|c|c|}\hline
L
&         8&        9&      10&       11&       12&Exact  value
                   \\[0pt]\hline
${\cal E}$(1)
&1.392376 &1.392357 &1.392344 &1.392349 &1.392351 &1.392352
                   \\[0pt]\hline
\end{tabular}

\vspace{0.3cm}
\noindent
For $L=8$ our estimate is closer by one order to the
exact value than the number $1.391655\pm 0.004562$ found in Ref.
\cite{21} on the basis of Wynn's $\ve$-algorithm.

The ground state energy $E(g)$ of the cubic anharmonic oscillator with
the Hamiltonian
$$H={1\over 2} (x^2+y^2)+ {g\over 4} [x^4+2 (1-\delta) x^2 y^2 +y^4]$$
depending of the anisotropy parameter $\delta$ and the value of the
coupling constant $g$ was estimated in Ref.\cite{15}. Those calculations
were based on the knowlege of the exact values of the asymptotic
parameters. The ground state energy estimated on the basis of our approach
proved to be very close to the values of Ref.\cite{15}.
The results given by two different methods are listed in Table II.
The stability of the result of processing $E(g)$ ($g/4=0.1$, 
$\delta= -2.5$)
with respect to the variation of the parameters $a$ and $b$ is
shown in Fig.2.

We have also studied the ground state energy $E(g)$ of the Yukawa 
potential $$V_g(x)=- {1\over x} \exp(-g x).$$
The dependence $E(g)$ presented in Fig.3 is in agreement with the
exact results \cite{21,22}. 
The exact critical value of $g$ when the bound state
disappears is $g_c=1.190612...$. Using Winn's $\ve$-algorithm 
\cite{21} gives $g_c=1.1836$. 
The best estimate for $g_c$ in the frame of our approach yields $g_c=1.191$.
In Fig.4 we demonstrate the behavior of
curves $E(b)$ for $g=1.11$ close to $g_c$ where
the domain of stability of the result of processing begins to 
dissipate (c.f. with Fig.2 for the cubic anharmonic oscillator 
where the parameter of expansion $g$ is small).

In the context of this report, it is interesting to study the convergence
of the RG series of certain conformal feild models, for which the exact
results are known, using our summation procedure. In two dimentions,
summation of $\ve$-series is a difficult problem because the parameter of
expantion is large ($\ve=2$). However, application of our approach
gives, as a whole, relatively good estimates. So, using five-loop
$\epsilon$-expansions for $O(N)$-symmetric model \cite{23} and setting
$N=2 \cos(\frac{\pi}{m})$, for the Ising model ($m=3$)
we obtain $\nu=0.925$, $\eta=0.220$, $\gm=1.650$ while
the exact theory predicts  $\nu=1$, $\eta=0.25$, $\gm=1.75$.
Obviously, the difference between our results and the exact values
does not exceed $8\%$. For the model describing polymers ($m=2$), 
our estimates are even better:  $\nu=0.747$ (exact $0.75$),
$\eta=0.190$ (exact $0.208$), and $\gm=1.352$ (exact $1.350$).
Above numbers are also in an accordance with the results of
Refs. \cite{8,10}, where a substantially different method 
was employed.

The analysis of the simple models fullfiled above enables one to apply
the summation approach introduced to finding numerical estimates of
important physical quantities in a number of real models. For
example, it was used for calculation of critical exponents for
the $N$-vector field models describing magnetic phase transitions
in cubic and tetragonal crystals \cite{19,24}. It can be expected that
the developed technique may be useful in such areas as QCD and QED as well.

We thank Professor D. I. Kazakov for interesting and
useful discussions of the problem of summation of the 
field-theoretical series during the XI International 
Conference "PROBLEMS OF QUANTUM FIELD THEORY".
We are grateful to Professor J. Zinn-Justin for important
remarks concerning the reference list of the paper.

\newpage

 TABLE II. Estimates of the ground state energy of the cubic anharmonic
 oscillator for various anisotropy parameters $\delta$, coupling 
 constant $g$, and approximation order $L$.

\vspace{0.5cm}
\hspace{0.2cm}
\begin{tabular}{|r|c|c|c|c|c|}
\multicolumn{6}{c}{Results by Kleinert {\em  et al.},  ${g/ 4} =0.1$  }
                   \\[6pt]\hline
$L\setminus \delta$&  -2,5    &  -1,5    &  -0,5    &   0,5    &   1,5
                   \\[0pt]\hline
     7  & 1,217107 & 1,192033 & 1,164803 & 1,134735 & 1,100604
                   \\[0pt]\hline
     9  & 1,217107 & 1,192034 & 1,164810 & 1,134736 & 1,100604
                   \\[0pt]\hline
    11  & 1,217107 & 1,192035 & 1,164810 & 1,134739 & 1,100604
                   \\[0pt]\hline
\multicolumn{6}{c}{}\\[-10pt]
\multicolumn{6}{c}{Our estimates, $g/4 =0.1$  }
                   \\[6pt]\hline
$L\setminus \delta$&-2,5  &  -1,5   &  -0,5   &   0,5   &   1,5
                   \\[0pt]\hline
               12 & 1,21705 & 1,19203 & 1,16480 & 1,134730 & 1,00600
                   \\[0pt]\hline
\multicolumn{6}{c}{}\\[-10pt]
\multicolumn{6}{l}{relative error $< 0.001 \%$ },\\
\multicolumn{6}{l}{relative deviation from the results of Ref.\cite{15}
$< 0.001\%$
}\\
\multicolumn{6}{l}{$0\leq b\leq 60 \quad 0.5\leq a\leq 1.5$}\\
\multicolumn{6}{c}{}\\[-10pt]
\multicolumn{6}{c}{Results by Kleinert {\em  et al.}, $g/4 =1.0$  }
                   \\[6pt]\hline
$L\setminus \delta$&  -2,5    &  -1,5    &  -0,5    &   0,5    &   1,5
                   \\[0pt]\hline
            7 & 1,941172 & 1,862806 & 1,773888 & 1,669172 & 1,535454
                           \\[0pt]\hline
            9 & 1,941172 & 1,862815 & 1,773909 & 1,669188 & 1,535425
                           \\[0pt]\hline
           11 & 1,941180 & 1,862823 & 1,773924 & 1,669199 & 1,535418
                   \\[0pt]\hline
\multicolumn{6}{c}{}\\[-10pt]
\multicolumn{6}{c}{Our estimates, $g/4 =1.0$  }
                   \\[6pt]\hline
$L\setminus \delta$&-2,5  &  -1,5   &  -0,5   &   0,5   &   1,5
                   \\[0pt]\hline
                12 & 1,9411 & 1,8627 &  1,7731 &  1,6691 &  1,5363
                       \\[0pt]\hline
\multicolumn{6}{c}{}\\[-10pt]
\multicolumn{6}{l}{relative error $< 0.06 \%$ }\\
\multicolumn{6}{l}{relative deviation from the results of Ref.\cite{15}
$< 0.05\%$
}\\
\multicolumn{6}{l}{$0\leq b\leq 60 \quad 0.5\leq a\leq 1.5$}
\end{tabular}

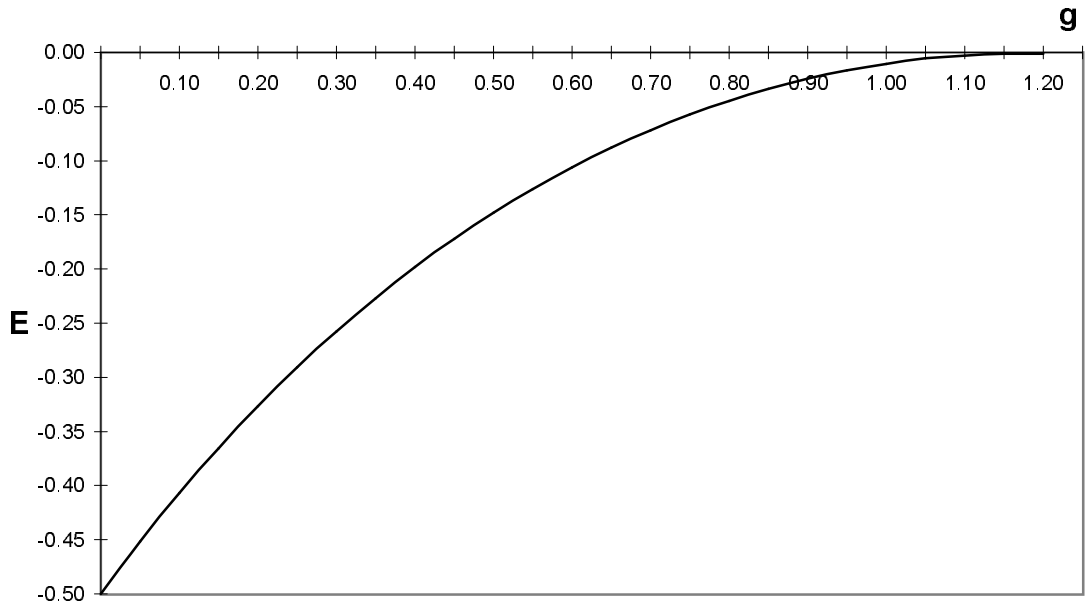
\end{document}